\documentclass[]{aastex631}
\usepackage[utf8]{inputenc}
\usepackage{graphicx}	
\usepackage{amsmath}	
\usepackage{amssymb}	

\newcommand{\VEC}[1]{\vec {#1} }

\begin{document}

%
\title{
 Evolutionary deformation toward the elastic limit
 by a magnetic field confined in the neutron--star crust 
}
\date{\today}

\author[0000-0002-4402-3568]{Yasufumi Kojima}
\author[0000-0002-2498-1937]{Shota Kisaka}
\affiliation{
Department of Physics, Graduate School of Advanced Science and Engineering,\\
Hiroshima University, Higashi-Hiroshima, Hiroshima 739-8526, Japan}
\author[0000-0002-9072-4744]{Kotaro Fujisawa}
\affiliation{
Department of Physics, Graduate School of Science, University of Tokyo, Bunkyo-ku, Tokyo 113-0033, Japan}

\begin{abstract}
Occasional energetic outbursts and anomalous X-ray luminosities
are expected to be powered by the strong magnetic field in a neutron star.
For a very strong magnetic field, elastic deformation becomes excessively large such that it leads to crustal failure.
We studied the evolutionary process driven by the Hall drift for a magnetic field confined inside the crust.
Assuming that the elastic force acts against the Lorentz force,
we examined the duration of the elastic regime and
maximum elastic energy stored before the critical state.
The breakup time was longer than that required for extending the field to the exterior,
because the tangential components of the Lorentz force vanished in the fragile surface region.
The conversion of large magnetic energy, confined to the interior, into Joule heat 
is considered to explain the power for central compact objects.
This process can function without reaching its elastic limit,
unless the magnetic energy exceeds $2\times 10^{47}$~erg, which
 requires an average field strength of $2\times10^{15}$~G.
Thus, the strong magnetic field hidden in the crust is unlikely to cause outbursts.
Furthermore, the magnetic field configuration can discriminate between central compact objects and magnetars.
\end{abstract}
%
%
\keywords{Neutron stars; Compact objects; Magnetars; High-energy astrophysics}
%

\section{Introduction}

\begin{figure}\begin{center}
\includegraphics[width=\columnwidth]{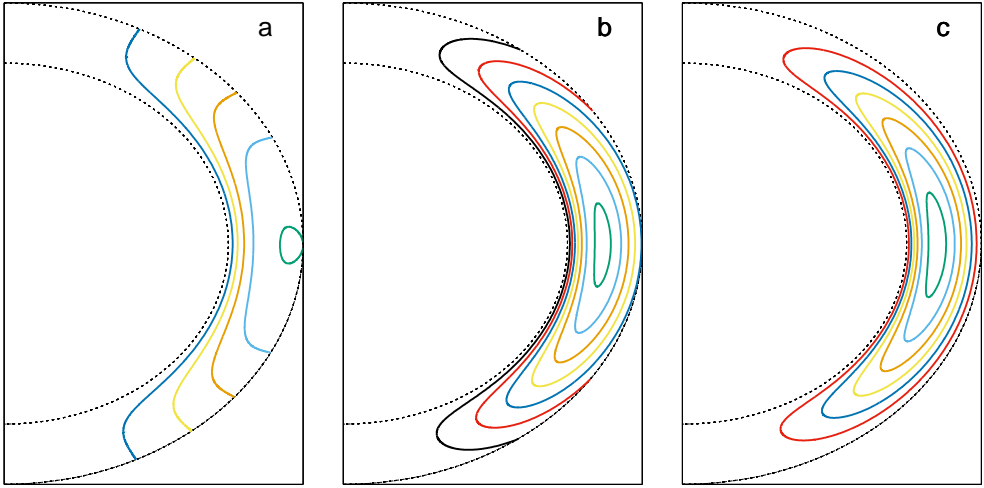}
\caption{ 
 \label{Fig.Bgeometry}
Magnetic field geometry in crust.
Contours of magnetic function $\Psi(r,\theta)$ 
for a poloidal field are demonstrated.
From the left to right panels,
the field is more confined, and the toroidal component is also involved in the interior.
}
\end{center}\end{figure}

%
The magnetic field strength on a neutron-star surface is typically approximately $10^{12}$~G.
However, there are two peculiar classes whose field strengths
significantly deviate from the average. They exhibit unusual activities, with
their energy considered to be supplied by an intense magnetic field.
Magnetars, except for a few sources, have a strong dipole field $\ge10^{14}$~G and
 exhibit energetic outbursts or flares.
The X-ray luminosity is very bright in the range 
$10^{32}$ --$10^{36}$ erg${\rm{s}}^{-1}$, which
exceeds the spin-down luminosity for most sources,
in contrast to normal pulsars \citep[][for review]
{2015RPPh...78k6901T,2017ARA&A..55..261K,2019RPPh...82j6901E,2021ASSL..461...97E}.
Central compact objects (CCOs) 
located at the centers of supernova remnants are X-ray sources with luminosity 
$\sim 10^{32}$ --$10^{34}$ erg${\rm{s}}^{-1}$.
A few CCOs show pulsations; hence, the magnetic dipole field is 
estimated to be $\sim 10^{10} -10^{11}$~G,
\citep{2013ApJ...765...58G}.
Their X-ray luminosities are comparable to those of quiescent magnetars
\citep[][]{2017ARA&A..55..261K},
and exceed the kinetic energy loss. 
To explain the X-ray luminosity, CCOs are considered to have an intense magnetic field
$\sim 10^{14}$~G inside neutron stars, although the surface field is weak.
Strong fields near the surface or inside the crust 
may explain the 
 nonuniform temperature of PSR J0822-4300 in Puppis A
\citep{2010ApJ...724.1316G}
and large pulse fraction of PSR J1852+0040 in Kes 79
\citep{2012ApJ...748..148S}.
Most of the magnetic field in CCOs is considered to be buried by the fallback of the supernova material
\citep{2011MNRAS.414.2567H,2012MNRAS.425.2487V}.
Numerical simulations can be used to solve the field geometry of the proto-neutron star
\citep[see, e.g.][for recent development]{2022MNRAS.516.1752M}.
Such a strong field $\sim 10^{14}$ G is crucial for studying magnetized neutron stars.
The Lorentz force is comparable in magnitude to the elastic force in the crust.
However, only static magneto-elastic equilibria of the crust have been studied 
thus far~\citep{2021MNRAS.506.3936K,2022MNRAS.511..480K,2023MNRAS.519.3776F}.
These studies demonstrated that neutron star models with
strong magnetic fields are possible owing to the elasticity of the crust.
A magnetic field configuration was assumed in these studies.
However, it is unclear whether a sufficient range of magnetic field was covered or not.
Herein, we consider the effect of the elastic force
in the equilibria of magnetized neutron stars from a different perspective.
In other words, we examined the process toward the elastic limit
according to magnetic field evolution.
Suppose that the magnetized crust settles in the force balance at a particular time.
However, the magnetic field is not fixed and evolves on a secular timescale.
Thus, elastic displacement is induced from the initial position 
to balance the Lorentz force according to field evolution.
Simultaneously, shear stress in the crust gradually accumulates 
and reaches a critical limit.
Beyond this threshold, the crust cracks~\citep[][for a seminal paper]{1992ApJ...392L...9D,1995MNRAS.275..255T}
or responds plastically. 
Two possibilities are discussed owing to the lack of 
a sufficient understanding of the material properties.
A sudden crust breaking can produce a magnetar outburst and/or a fast radio 
burst~\citep{2016ApJ...833..189L,2018MNRAS.480.5511B,2019MNRAS.488.5887S}.
Second, plastic flow beyond a critical point is crucial for the long-term evolution
\citep{2016ApJ...824L..21L},
and a coupled system between the flow and the magnetic field was numerically solved
\citep{2019MNRAS.486.4130L,2020MNRAS.494.3790K,2021MNRAS.506.3578G}.
Therefore, it is important to explore the timescale up to the critical limit, 
and the elastic energy deposited during the evolution.
In this study, we assumed that the stellar structure is always barotropic.
The initial state of evolution is described by 
 magneto-hydrodynamic (MHD) equilibrium without the elastic force. 
 This equilibrium does not involve electrons.
\citep[][]{2013MNRAS.434.2480G,2014MNRAS.438.1618G};
thus, the magnetic field tends toward Hall equilibrium on a secular timescale.
The Hall--drift timescale, an important indicator in the evolution, becomes
shorter as the magnetic field strength increases.
Thus, this study is relevant to neutron star crusts with strong magnetic fields.
A previous study considered the evolution of a magnetic field that extended from the crust to the exterior
\citep[][referred to as Paper I]{2022ApJ...938...91K}.
However, the toroidal component of the magnetic field was ignored.
It is important to examine the evolution of different magnetic field geometries.
The broad classification is based on whether the field is confined inside the crust 
or spreads out to the magnetosphere.
This possibility is schematically illustrated in Figure~\ref{Fig.Bgeometry}.
For simplicity, we assume that the field is purely dipolar and is expelled 
from the neutron--star core.
The extended case shown in Figure~\ref{Fig.Bgeometry}-a (left panel)
 corresponds to the magnetar model considered in Paper I.
 The toroidal component of the magnetic field cannot emerge in an exterior vacuum; it is confined inside a loop in the meridional plane for the poloidal component, 
as shown in Figure~\ref{Fig.Bgeometry}-b (middle panel) 
and \ref{Fig.Bgeometry}-c (right panel).
The toroidal magnetic energy potentially increases as the loop region expands. 
In contrast to Paper I, this study considered the entirely confined case, shown in 
Figure~\ref{Fig.Bgeometry}-c.
Both the toroidal and poloidal components are confined in the crust.
The magnetic field geometry can be applied to that of CCOs.
The remainder of this paper is organized as follows. The models and equations used in this study are discussed in Section 2. 
We calculated the quasi-stationary evolution of the shear strain induced by the Hall drift of a magnetic field.
We estimated the critical time beyond which the elastic equilibrium was no longer possible as well as the
elastic energy during evolution.
The numerical results are presented in Section 3.
Finally, our conclusions are presented in Section 4.

\section{Mathematical formulation}

\subsection{Barotropic Equilibrium in Crust}

\begin{figure}\begin{center}
  \includegraphics[scale=0.8]{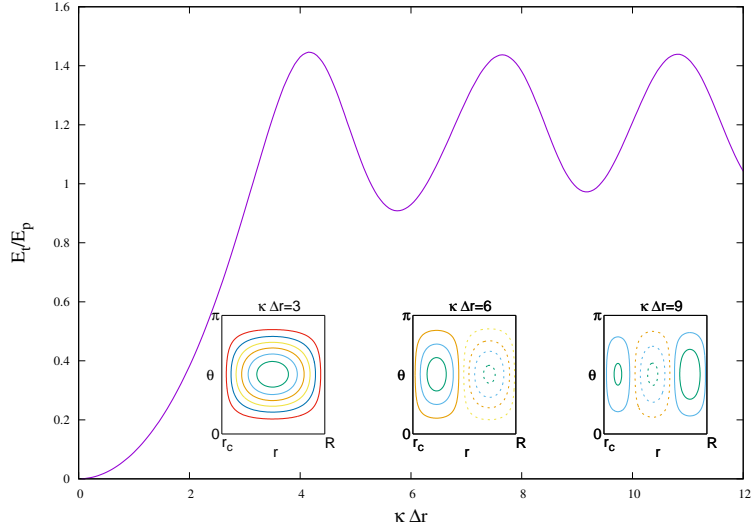}%
\caption{ 
\label{Fig.Evsk}
Energy ratio as a function of $\kappa\Delta{r}$.
Contours of $\Psi$ in the $r - \theta$ plane for $\kappa\Delta{r}=$3, 6, and 9 are shown.
Contours with negative values of $\Psi$ are indicated by dotted lines. 
}
\end{center}\end{figure}

Our consideration was limited to the inner crust of a neutron star,
where the mass density ranges from 
$\rho_{c}=1.4\times 10^{14}$ g cm$^{-3}$ at the core--crust boundary $r_{c}$ 
to the neutron drip density $\rho_{1}=4\times 10^{11}$ g cm$^{-3}$ at $R=12$ km.
We ignored the outer crust and treated the exterior region of $r>R$ as a vacuum.
The crust thickness $\Delta{r}\equiv R-r_{c}$ was assumed to be $\Delta{r}/R=0.05$ and $\Delta{r}=0.6$ km.
The spatial density profile in $r_c\le r \le R$ is approximated as~\citep{2019MNRAS.486.4130L}
\begin{equation}
{\hat\rho}\equiv\frac{\rho}{\rho_{c}}
=\left[1-\left(1-\left(\frac{\rho_1}{\rho_{c}}\right)^{1/2}
 \right)\left(\frac{r-r_{c}}{\Delta{r}}\right)\right]^2.
\label{dnsprofile.eqn}
 \end{equation}
We consider the equilibrium in the crust. Under the Newtonian approximation,
static force balance between the pressure $P$, gravity, and Lorentz force is expressed as
\begin{equation}
-\frac{1}{\rho}{\VEC {\nabla}}P-{\VEC {\nabla}}\Phi_{\rm g}
+\frac{1}{c\rho} {\VEC {j}}\times {\VEC {B}}=0,
  \label{Forcebalance.eqn}
\end{equation}
where $\Phi_{\rm g}$ is the gravitational potential including the centrifugal term.
We assume a barotropic distribution $P=P(\rho)$,
and the sum of the first two terms in Equation~(\ref{Forcebalance.eqn}),
is expressed as $ -{\VEC \nabla}\Phi_{\rm eff}$.
The third term has a magnitude $\sim 10^{-7}(B/10^{14}{\rm G})^2$ times
smaller than those for the first and second terms.
The deviation due to the Lorentz force is sufficiently small to 
be treated as a perturbation of the background equilibrium.
We assumed an axial symmetry for the magnetic-field configuration.
The poloidal and toroidal components of the magnetic field are expressed by two functions, $\Psi$ and $S$, respectively, as follows:
\begin{equation}
{\VEC{B}}={\VEC{\nabla}}\times \left(\frac{\Psi}{\varpi}{\VEC{e}}_{\phi}\right)
+\frac{S}{\varpi}{\VEC{e}}_{\phi},
    \label{eqnDefBB}
\end{equation}
where $\varpi=r\sin\theta$ is the cylindrical radius and
$\VEC{e}_{\phi}$ is the azimuthal unit vector in $(r,\theta, \phi)$ coordinates.
For barotropic equilibrium, the current function $S$ should be a function of $\Psi$, and 
the azimuthal component $j_{\phi}$ of the electric current is described in the form 
\citep[e.g.,][]{2005MNRAS.359.1117T}
\begin{equation}
\frac{4\pi j_{\phi}}{c} = \rho \varpi\frac{dK}{d\Psi}
+\frac{S}{\varpi}\frac{dS}{d\Psi},
    \label{MHDeqil.eqn}
\end{equation}
where $K$ denotes a function of $\Psi$.
Further, the acceleration ${\VEC{a}}$ owing to the Lorentz force is reduced to
\begin{equation}
 {\VEC{a}}\equiv\frac{1}{c\rho}{\VEC{j}}\times{\VEC{B}}
=\frac{1}{4\pi}{\VEC{\nabla}}K.
    \label{EMacc0.eqn}
\end{equation}
Thus, the force balance in Equation~(\ref{Forcebalance.eqn}) is
described by the gradient terms of scalar functions.
We adopted a simple linear function of $\Psi$ for $K(\Psi)$ and $S(\Psi)$.
$K=K_{0}\Psi$ and $S=\kappa\Psi$, where $K_{0}$ and $\kappa$ are constant.
For the dipole field, function $\Psi$ is expressed using 
the Legendre polynomial of $l=1$, that is, $\Psi=g(r)\sin^2\theta$. 
After the decomposition of the angular part,
the azimuthal component of the Amp{'e}re law is reduced to
\begin{equation}
g^{\prime\prime}-\frac{2g}{r^2}+\kappa^2 g=-K_{0}\rho r^2,
    \label{g1diff.eqn}
\end{equation}
where prime $^{\prime}$ denotes a derivative with respect to $r$.
We consider a magnetic field confined in the crust such that
the radial function $g$ is obtained by solving Equation~(\ref{g1diff.eqn}) using
 boundary conditions $g(r_c)=g(R)=0$.
The solutions for Equation~(\ref{g1diff.eqn}) without the source term are obtained using spherical Bessel functions. 
A homogeneous solution satisfies the following boundary conditions:
only for specific values,
  $\kappa\Delta{r}\approx n \pi~(n=1,2,\cdots)$.
  This solution corresponds to a force-free case ${\VEC{j}}\times{\VEC{B}}=0$, 
  that is, $K_{0}=0$ in Equation~(\ref{EMacc0.eqn}).
The constant $K_{0}$ determines the overall magnetic field strength, whereas
 $\kappa$ determines the ratio of the poloidal and toroidal components.
The dipolar magnetic field considered in Paper I 
is purely poloidal ($\kappa=0$) and extends to the exterior vacuum.
The magnetic energy $E_{\rm{p}}$ stored inside the crust
is expressed as $E_{\rm{p}}=3.8B_{0}^2R^3$, using
field strength $B_{0}$ at the surface.
When studying different models confined to the crust,
we always fixed the poloidal magnetic energy as $E_{\rm{p}}=3.8B_{0}^2R^3$. The average strength is approximated as $12B_{0}$, whereas the normalization $B_{0}$ also fixes $K_{0}$ for each model. 
Figure~\ref{Fig.Evsk} shows the energy ratio
of toroidal components $E_{\rm{t}}$ to poloidal components $E_{\rm{p}}$
 as functions of $\kappa\Delta{r}$.
 A similar result was obtained for the magnetic field confined in 
 the whole star \citep[][]{2013MNRAS.432.1245F}.
The ratio increased with $\kappa \Delta{r}$ and reached a maximum $\sim 1.4$.
 With further increase, the ratio
 oscillated between the minimum $\sim 1$ and the maximum $\sim 1.4$.
Further, the spatial structure of the magnetic function $\Psi$ 
changed continuously; the radial nodes increased with $\kappa\Delta{r}$
as shown in Figure ~\ref{Fig.Evsk}.
Node number $n$ is approximated as $n\approx\kappa\Delta{r}/\pi$.
%

   \subsection{Magnetic--Field Evolution}

The force balance expressed as Equation~(\ref{Forcebalance.eqn})
is not fixed on a secular timescale because
the Lorentz force ${\VEC{j}}\times{\VEC{B}}$ gradually
changes owing to magnetic field evolution. 
The evolution of the crustal magnetic field was governed by the following induction equation:
\begin{align}
\frac{\partial}{\partial t}{\VEC{B}}=-{\VEC{\nabla}}\times
\left[\frac{1}{e n_{\rm{e}}}{\VEC{j}}\times{\VEC{B}}+
\frac{c}{\sigma_{\rm{e}}}{\VEC{j}}\right],
  \label{Frad.eqn}
\end{align}
where $n_{\rm{e}}$ is the electron number density and $\sigma_{\rm{e}}$ is the electrical conductivity.
The first term in Equation~(\ref{Frad.eqn}) represents the Hall drift,
and the second term represents the magnetic decay due to ohmic dissipation.
The timescales associated with these processes are estimated as
\begin{align}
& \tau_{\rm{H}} =\frac{4\pi e n_{\rm{ec}} (\Delta{r})^2}{c B_{0}}
=7.9\times 10^5
\left(\frac{ B_{0} }{10^{14}{\rm{G}} }\right)^{-1}{\rm{yr}}, 
   \label{timescale.hall}
\\
& \tau_{\rm Ohm}=\frac{4\pi \sigma_{\rm{ec}} (\Delta{r})^2}{c^2}
  =2.1\times 10^6{\rm yr},
   \label{timescale.ohm}
\end{align}
where $B_{0}$ denotes the normalization of the magnetic field 
strength, and crust thickness $\Delta{r}=0.05R=0.6$km
is used.
In Equations~(\ref{timescale.hall}) and (\ref{timescale.ohm}), we used the maximum values for $n_{\rm{e}}$ and $\sigma_{\rm{e}}$, 
that is, the values at the core--crust boundary. 
 \begin{equation}
n_{\rm{ec}}=3.4\times 10^{36}{\rm{cm}}^{-3}, 
~~~
\sigma_{\rm{ec}}=1.5\times 10^{24}{\rm{s}}^{-1}. 
\label{nuc.coefficient}
\end{equation}
The actual timescales may be smaller than those obtained using Equations~(\ref{timescale.hall}) and (\ref{timescale.ohm}), 
owing to the spatial dependence of $n_{\rm e}$, $\sigma_{\rm{e}}$,${\VEC{B}}$, and ${\VEC{j}}$.
Here, we estimate the magnetic decay for the confined 
models considered in the previous subsection, using numerical calculations.
The energy decay time $t_{\rm{d}}$ is defined as
\begin{equation}
  t_{\rm{d}}\equiv 
2E_{\rm{B}}\left( \int \frac{j^2}{\sigma_{\rm{e}}} dV \right)^{-1},
\label{tdiss.eqn}
\end{equation}
where $E_{\rm{B}}$ denotes the total magnetic energy.
$E_{\rm{B}}= E_{\rm{p}}+ E_{\rm{t}}$,
The field strength $B$ decreased by approximately 
$B \propto \exp(-t/t_{\rm{d}})$. Further,
the magnetic energy 
($E_{\rm{B}}\propto B^2 \propto \exp(-2t/t_{\rm{d}})$) was dissipated
by one order in the period 
$(0.5\ln10) \times t_{\rm{d}}  \approx 1.2 t_{\rm{d}}$.
We calculated $t_{\rm{d}}$ 
using an analytical model for the electric conductivity distribution
\citep{2019MNRAS.486.4130L};
\begin{equation}
n_{\rm e}  =n_{\rm e c}\left( 0.44 {\hat \rho}^{2/3} + 0.56 {\hat \rho}^{2} \right), 
~~~
\sigma_{\rm{e}}=\sigma_{\rm{ec}}
\left( \frac{n_{\rm e}}{n_{\rm e c}} \right)^{2/3},
%
 \end{equation}
where ${\hat \rho}=\rho/\rho_{c}$ is expressed as Equation~(\ref{dnsprofile.eqn}).

\begin{figure}\begin{center}
\includegraphics[scale=0.8]{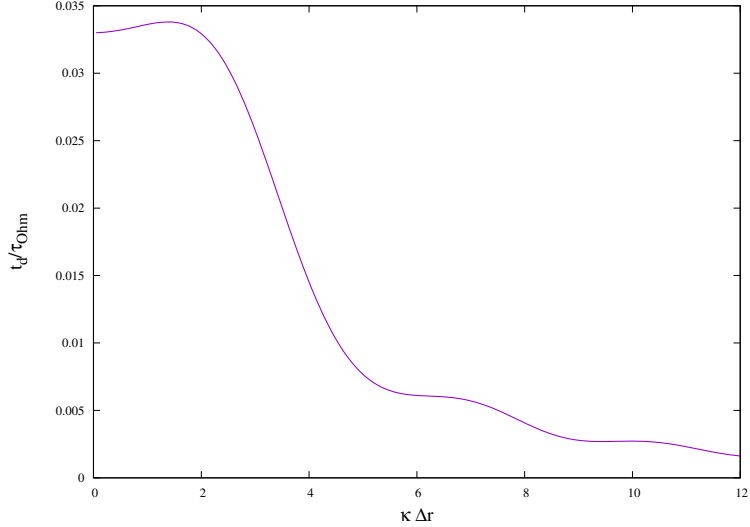}%
\caption{ 
 \label{Fig.rttdto}
Magnetic decay time $t_{\rm{d}}/\tau_{\rm{Ohm}}$ as a function of $\kappa\Delta{r}$.
}
\end{center}\end{figure}

\begin{figure}\begin{center}
\includegraphics[width=\columnwidth]{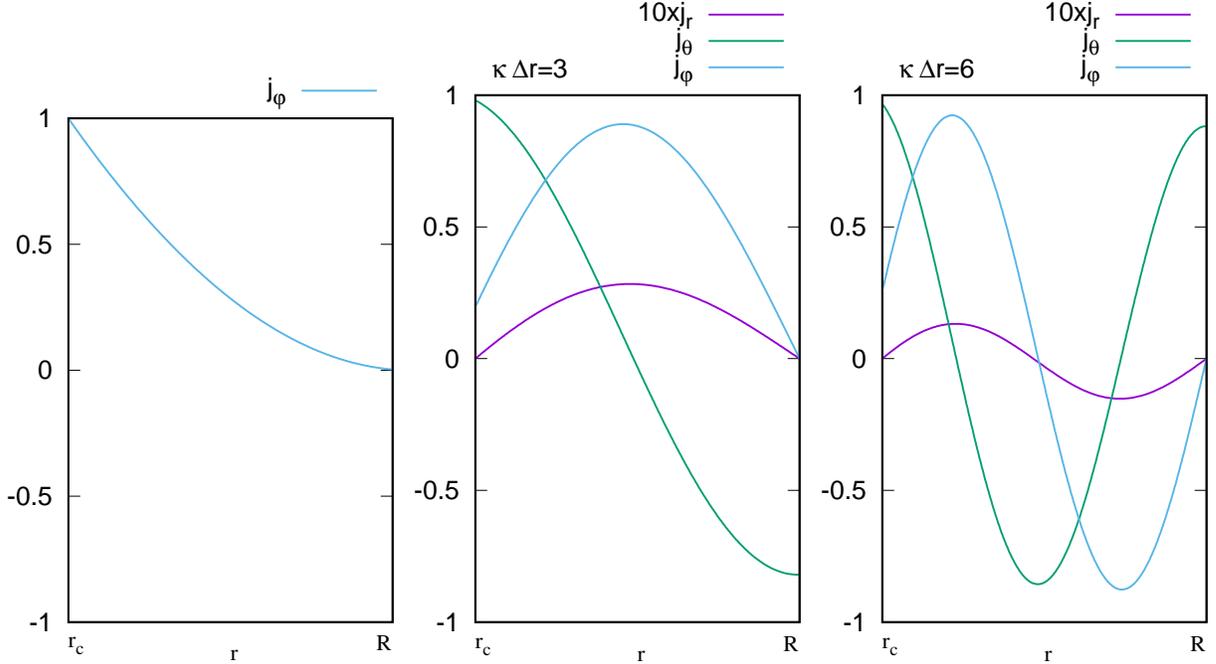}%
\caption{ 
 \label{Fig.current}
Electric currents in crust.
Radial functions $(j_{r}/\cos\theta, j_{\theta}/\sin\theta, j_{\phi}/\sin\theta)$
are shown for three models.
The left to right panels correspond to
the extending field, confined field with $\kappa \Delta r=3$,
and confined field with $\kappa \Delta r=6$, respectively.
The amplitudes are normalized by the maximum of $|{\VEC{j}}|$.
The radial component $j_{r}$ for the confined models is small, typically 
$j_{r}\sim (\Delta r/R)\times j_{\theta}=j_{\theta}/20$,
such that $j_{r}$ in the figures is multiplied by a factor of 10.
}
\end{center}\end{figure}

Figure~\ref{Fig.rttdto} shows the results for a confined field.
The small ratio $t_{\rm{d}}/\tau_{\rm{Ohm}} \sim 10^{-2}$ originates largely from the
choice of the maximum value of electric conductivity $\sigma_{\rm{ec}}$ in Equation~(\ref{timescale.ohm}).
Thus, we found that the dissipation time is of the following order: 
$3\times(10^{-3}-10^{-2})\times \tau_{\rm{Ohm}}$
$\sim 6\times 10^{3} - 6\times 10^{4}$ yr.
During this period, the magnetic energy was converted to heat at a rate of 
\begin{equation}
{\dot{Q}} =\frac{2E_{\rm{B}}}{t_{\rm{d}}}
\approx 7 \times (10^{34} -10^{35} )\times
 \left(\frac{B_{0}}{10^{14}{\rm{G}}} \right)^{2}
{\rm{erg~s}}^{-1}.
\end{equation}
This power is sufficient to supply the X-ray luminosity of CCO
$\sim 1.5\times 10^{32} - 10^{34}$ erg s$^{-1}$.
To study the effect of the magnetic geometry, we also compared 
 the dissipation  timescale for extending the field to the exterior, 
where $t_{\rm{d}}/\tau_{\rm{Ohm}}=5.9\times 10^{-2}$.
This value was larger than that for the confined field.
Among the confined field models, $t_{\rm{d}}/\tau_{\rm{Ohm}}$ decreased slightly with an increase in $\kappa\Delta{r}$.
As the radial number increased, the typical length and the dissipation time decreased.
The stepwise curve of $t_{\rm{d}}/\tau_{\rm{Ohm}}$ in Figure~\ref{Fig.rttdto} corresponds to this transition.
We considered the electric current distribution.
The angular dependence is expressed as
 $j_{r}\propto \cos\theta$,
 $j_{\theta} \propto \sin\theta$, and $j_{\phi} \propto\sin\theta$
 because the magnetic field is dipolar ($l=1$). 
Figure~\ref{Fig.current} shows the radial functions of the electric currents for the three models.
The current for the extended model in the left panel is the maximum at the inner boundary $r=r_{c}$ and decreases radially.
However, for the confined models, we found that $j_{\theta}$ is large at the
surface, while $j_{r}=j_{\phi}=0$.
Component $j_{\theta}\ne 0$
 near the surface decayed significantly because of the low 
electric conductivity, thereby resulting in a larger value of $t_{\rm{d}}/\tau_{\rm{Ohm}}$.
%

   \subsection{Hall-drift Evolution}

The Hall-Ohmic equation (\ref{Frad.eqn}) was studied 
for axially symmetric models~\citep{2007A&A...470..303P,2012MNRAS.421.2722K,2013MNRAS.434..123V,
2013MNRAS.434.2480G,2014MNRAS.438.1618G,2021CoPhC.26508001V},
and for 3D models 
\citep{2015PhRvL.114s1101W,2019CoPhC.237..168V,2020ApJ...903...40D,
2020MNRAS.495.1692G,2021ApJ...909..101I}.
Here, we limit the evolution to the early phase such that 
the calculation is simplified.
By comparing the two timescales expressed as Equations~(\ref{timescale.hall}) and (\ref{timescale.ohm}),
it was revealed that the magnetic field evolution is governed by Hall drift in the strong-magnetic-field regime.
In the range $t < t_{\rm{d}} \sim 10^4$ y, 
we may neglect Ohmic decay in Equation~(\ref{Frad.eqn}) and 
the induction equation is reduced to
\begin{equation}
\frac{\partial}{\partial t}{\VEC{B}}=-{\VEC{\nabla}}\times
\left[
\frac{1}{e n_{\rm{e}}}{\VEC{j}}\times {\VEC{B}}
\right]
= -{\VEC{\nabla}}\chi\times{\VEC{a}}
 - \chi{\VEC{\nabla}}\times{\VEC{a}},
 \label{BHevl.eqn}
\end{equation}
where
\begin{equation}
\chi \equiv \frac{c\rho}{e n_{e}}
= \frac{4\pi \rho_{c} (\Delta r)^2}{\tau_{\rm H}B_{0}}{\hat \chi}.
\label{chidef.eqn}
\end{equation}
In Equation~(\ref{chidef.eqn}),
${\hat{\chi}}$ is the non-dimensional ratio
of mass density to electron number density, and
is approximated by a smooth analytic function to fit the data given by 
\citep[][Paper I]{2001A&A...380..151D},
\begin{equation}
{\hat{\chi}}=\frac{ {\hat{\rho}}^{1/2}}{0.32+0.66  {\hat{\rho}}},
 \label{chiftfm.eqn}
\end{equation}
where ${\hat \rho}=\rho/\rho_{c}$ is expressed as Equation~(\ref{dnsprofile.eqn}).
The second term in Equation~(\ref{BHevl.eqn})
vanishes at $t=0$ owing to Equation~(\ref{EMacc0.eqn}), because
barotropic equilibrium is assumed.
Moreover, the first term vanishes when $\chi$ is constant.
In other words, the barotropic MHD equilibrium is the Hall equilibrium for electrons \citep{2013MNRAS.434.2480G}.
We consider the magnetic-field evolution driven by
nonuniform distribution of $\chi$.
The early phase of barotropic MHD equilibrium is governed by 
\begin{equation}
\frac{\partial}{\partial t}{\VEC{B}}=
-\chi^{\prime} a_{\theta} {\VEC{e}}_{\phi}.
%
\end{equation}
The azimuthal component, $B_{\phi}$, changes linearly with time, $t$.
We ignore the changes in the poloidal magnetic field $\delta {\VEC{B}}_{p}$, and 
 in the relevant azimuthal current $\delta j_{\phi}$.
  The early phase of the toroidal magnetic field is expressed as
  $\delta B_{\phi}$ and can be approximated as
\begin{equation}
 \delta B_{\phi} =
\frac{\delta S}{\varpi}\left(\frac{t}{\tau_{\rm H}}\right),
  \label{delB31.eqn}
\end{equation}
where the function $\delta S(r,\theta)$ is explicitly expressed in terms of
the Legendre polynomial $P_{l}(\cos\theta)$ of $l=2$ as
\begin{equation} 
\delta S\equiv y(r) \sin\theta P_{2,\theta}
=\frac{2K_{0}\rho_{c}({\Delta}{r})^2}{3B_{0}} 
{\hat{\chi}}^{\prime} g \sin \theta P_{2,\theta},
   \label{barsfn.eqn} 
\end{equation}
where the radial function, $y$ is defined for convenience.
The poloidal current changes are associated with $\delta B_{\phi}$; thus, the Lorentz force,
$\delta {\VEC f}=c^{-1}(\delta{\VEC j} \times {\VEC B}+
{\VEC j} \times \delta{\VEC B})$ also changes and is
explicitly written as 
\begin{equation}
    \delta {\VEC f}=-[{\VEC \nabla}(S\delta S) 
  + {\VEC \nabla} \Psi \times {\VEC \nabla}\delta S]  
   \left(\frac{t}{4\pi \varpi^2 \tau_{\rm H}}\right).
    \label{delEMforce.eqn}
\end{equation}
%

   \subsection{Quasi-stationary Elastic Response}

The force balance deviates slightly from the initial state owing to
the change in the Lorentz force through magnetic field evolution.
The acceleration associated with Equation~(\ref{delEMforce.eqn}) is generally the sum of the
solenoidal and irrotational components.
The solenoidal part of the Lorentz force should be
balanced by additional forces when the material distribution is barotropic;
that is, the sum of the pressure and gravitational potential terms is
expressed as $-{\VEC{\nabla}}\delta \Phi_{\rm{eff}}$.
The elastic force $\delta {\VEC{h}}$ in 
the solid crust is assumed to act against the solenoidal part.
Note that the force is purely solenoidal for incompressible motion
in the case of a constant shear modulus $\mu$; that is, 
$\delta {\VEC h} = -\mu {\VEC{\nabla}}\times {\VEC{\nabla}}\times {\VEC{\xi}}$
with the displacement vector ${\VEC{\xi}}$
~\citep[for example][]{1959thel.book.....L}. 
In general, the force contains both solenoidal and irrotational parts,
and is expressed by the trace-free strain tensors $\sigma_{ij}$ and $\mu$.
\begin{equation}
   \delta h^{i} = {\nabla}_{j} \left(2\mu \sigma^{ij} \right)
\end{equation}
and
\begin{equation}
   \sigma_{ij} =\frac{1}{2}({\nabla}_{i} \xi_{j} +{\nabla}_{j} \xi_{i}),
\end{equation}
where the incompressible displacement is assumed as
${\VEC{\nabla}}\cdot {\VEC{\xi}} =0$. 
In addition, we assumed that the shear modulus $\mu$
is proportional to the density~\citep[Figure~43 in ][]{2008LRR....11...10C},
such that the shear speed $v_{s}$
 is constant throughout the crust and $\mu=v_{s}^2 \rho$, where
\begin{equation}
v_{s}=8.5 \times 10^{7} {\rm cm~s}^{-1}. 
%
\end{equation}
The shear modulus $\mu$ is the maximum,
$\mu_{c} \approx 10^{30} {\rm ~erg~cm^{-3}}$ at the core--crust interface
while it decreases toward the stellar surface, 
$\mu_{1} \approx 3\times 10^{27} {\rm ~erg~cm^{-3}}$ at $\rho_1$.
The elastic evolution was excessively slow; hence, 
the acceleration $\partial^2 \xi_{i}/\partial t^2$ can be ignored.
Consequently, the elastic force is balanced by the change in the Lorentz force
at any time, that is, quasi-stationary evolution.
Under the approximation that the solenoidal part, that is,
a ”curl” of acceleration owing to the Lorentz force 
should be balanced with that of the elastic force,
a set of equations is expressed as
\begin{equation}
( \delta {\VEC f}+ \delta {\VEC h})_{\phi}=0, 
\label{blanceeqn3a}
\end{equation}
\begin{equation}
   [{\VEC{\nabla}}\times \rho^{-1}(\delta{\VEC f}+\delta{\VEC h})]_{\phi}=0.
\label{rotforce3} 
\end{equation}
Note that we consider the azimuthal component only in Equation~(\ref{rotforce3}) because other poloidal components are redundant
when using Equation~(\ref{blanceeqn3a}) 
 and an axial symmetry ($\partial_{\phi}=0$).
The terms involving the Lorentz force in Equations~(\ref{blanceeqn3a}) and (\ref{rotforce3})
are expressed in terms of Legendre polynomials $P_{l}(\cos\theta)$ of $l=1,3$\footnote{
Initial barotropic equilibrium models are magnetically deformed with an ellipticity (deformation of $l=2$) 
$\sim 10^{-8} (B_{0}/(10^{14}{\rm{G}}))^{2}$~\citep[e.g.,][]{2021MNRAS.506.3936K}.
The quadrupole deformation does not change because the induced components are $l=1$ and 3.
}.
\begin{equation}
\delta f_{\phi}= -\sum_{l=1,3}  r^{-3} a_{l} P_{l,\theta}
 \left(\frac{t}{\tau_{\rm H}}\right).  
 \label{expndf3.eqn}
\end{equation}
\begin{equation}
[{\VEC{\nabla}}\times \rho^{-1}\delta{\VEC{f}} ]_{\phi}
= -\sum_{l= 1, 3}  r^{-1} b_{l} P_{l,\theta}
 \left(\frac{t}{\tau_{\rm H}}\right). 
 \label{expndfp.eqn}
\end{equation}
where $a_{l}$ and $b_{l}$ $(l=1,3)$ are radial functions expressed as
\begin{align}
& a_{1}=-\frac{3}{10\pi}(gy)^{\prime},
~~
a_{3}=-\frac{1}{10\pi}(2(gy)^{\prime}-5g^{\prime}y),
%
\\
&b_{1}=\frac{3\kappa}{10\pi}
\frac{(gy)^{\prime}}{\rho r^2},
~~
b_{3}=\frac{\kappa}{10\pi}\left(\frac{2(gy)^{\prime}}{\rho r^2}
+5\left( \frac{1}{\rho r^2}\right)^{\prime}gy
\right).
\label{cfb1and3.eqn}
\end{align}
The elastic displacement growth with $t$ is explicitly expressed as
\begin{equation}
(\xi_{r}, \xi_{\theta}, \xi_{\phi})=
 \sum_{l=1,3}
\left(\frac{l(l+1)x_{l}}{r^2}P_{l},
~\frac{x_{l}^{\prime}}{r}P_{l, \theta},
~-rk_{l}P_{l, \theta}
\right)\left(\frac{t}{\tau_{\rm H}}\right),
\end{equation}
where the radial functions $x_{l}$ and $k_{l}$ $(l=1,3)$ are determined using
Equations~(\ref{blanceeqn3a}) and (\ref{rotforce3})
~\cite[][]{2022MNRAS.511..480K};
\begin{equation}
 (\mu r^{4} k_{l}^{\prime})^{\prime}
    -(l-1)(l+2)\mu r^{2} k_{l}=-a_{l},
    \label{klexpd.eqn}    
\end{equation}
\begin{equation}
    x_{l}^{\prime\prime}
    -\frac{l(l+1)}{r^2}x_{l}+\frac{1}{\mu}w_{l}=0,
    \label{flexpd.eqn}
\end{equation}
\begin{equation}
\left[\rho^{-1}w_{l}^{\prime}
+2\left(\frac{\mu^{\prime}}{\rho r}\right) x_{l}^{\prime}
\right]^{\prime}
-\frac{l(l+1)}{r^2}
\left[\rho^{-1}w_{l}^{\prime}
+2\left(\frac{\mu^{\prime}}{\rho}\right)^{\prime}x_{l}
\right] =- b_{l}.
    \label{glexpd.eqn}
\end{equation}
We now discuss the boundary conditions for a set of ordinary differential equations 
(\ref{klexpd.eqn}) - (\ref{glexpd.eqn}).
Across these surfaces at $r_c$ or $R$, the change in the total stress--tensor
should vanish for force balance.
In other words, $2\mu \sigma_{ri}+\delta T_{ri} =0~(i=r,\theta, \phi)$,
where $\delta T_{ij}$ denotes magnetic stress.
Owing to the fact that $\delta {\VEC{B}}_{p}=0$ and $B_{r}=B_{\phi}=0$ at the boundaries,
 the boundary conditions are reduced to
$\sigma_{r r}= \sigma_{r \theta}=\sigma_{r \phi}=0$. 
These conditions for the radial functions $k_{l}$ 
$x_{l}$ and $w_{l}$ $(l=1,3)$ at $r_{c}$ and $R$ can be written explicitly as
\begin{align} 
& rx_{l}^{\prime}-2x_{l}=0,
   \label{bcT11}
 \\
&
2 \mu r x_{l} ^{\prime}-2\mu l(l+1) x_{l} 
 +r^{2} w_{l} =0,
\label{bcT12}
 \\
&  k_{l}^{\prime} =0.
  \label{bcT13}
\end{align}

\section{Results}

\subsection{Breakup Time and Accumulated Energy}

\begin{figure}\begin{center}
  \includegraphics[scale=0.8]{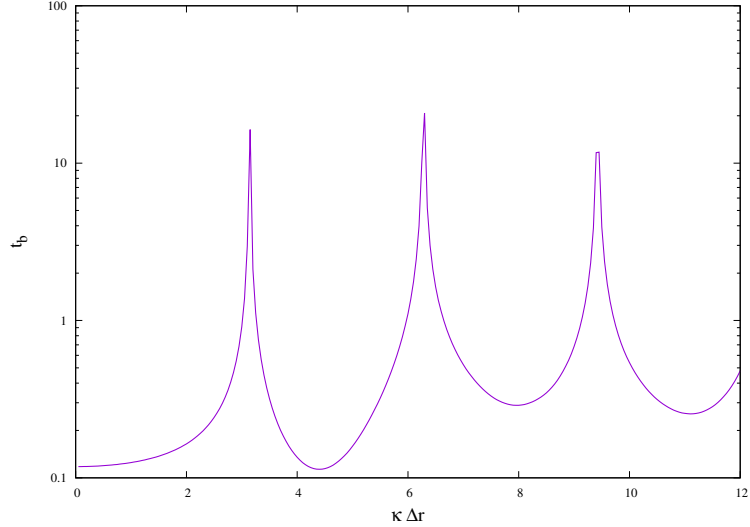}%
\caption{ 
 \label{Fig.tbvsk}
Breakup-time ${\hat{t}}_{b}$ as a function of $\kappa \Delta{r}$.
}
\end{center}\end{figure}

By solving the differential equations, we obtain the shear stress $\sigma_{ij}$
whose magnitude increases with time, whereas the spatial profile remains unchanged.
The numerical calculations provided the maximum shear stress
with respect to $(r,\theta)$ in the crust.
Elastic equilibrium can be achieved until the breakup time $t_{*}$, 
only when the shear strain satisfies a particular criterion.
We adopted the following (von Mises criterion) to determine the elastic limit:
\begin{equation}
\frac{1}{2}\sigma_{ij}\sigma^{ij}
\le (\sigma_{\rm{max}})^{2},
%
   \label{criterion}
\end{equation}
where $\sigma_{\rm{max}}$ denotes the number of
$\sigma_{\rm{max}} \approx 10^{-2}-10^{-1}$
~\citep{2009PhRvL.102s1102H,2018PhRvL.121m2701C,2018MNRAS.480.5511B}. 
Thus, the period of the elastic response is expressed as 
\begin{align}
 \label{limitVA.eqn}
    t \le t_{*} & \equiv 
    {\hat{t}}_{b}\left(\frac{ \sigma_{\rm{max}}}{0.1}\right)\left(\frac{v_{s}}{v_{a}} \right)^2 \tau_{{\rm H}}
\\
&= 2.8 \times 10^{6} \times {\hat{t}}_{b} 
\left(\frac{\sigma_{\rm{max}}}{0.1}\right)
 \left(\frac{B_{0} }{10^{14}{\rm G}}\right)^{-3}{\rm{yr}},
%
\end{align}
where ${\hat{t}}_{b}$ is a numerical factor that depends on parameter $\kappa$.
The criterion in Equation (\ref{limitVA.eqn}) depends on the ratio of shear to magnetic forces, and 
 is characterized by the shear speed $v_{s}$ 
and {Alfv{\'e}n} speed $v_{a}$, which is defined as
\begin{equation}
    v_{a}=\frac{B_0}{\sqrt{4\pi\rho_{1}}}
=4.5\times 10^{7}
\left(\frac{B_{0}}{10^{14}{\rm{G}}}
\right) {\rm{cm}~s}^{-1}, 
\end{equation}
where $B_{0}$ is determined by the poloidal magnetic energy $E_{\rm{p}}$
as discussed in Subsection 2.1. 
Figure \ref{Fig.tbvsk} illustrates ${\hat{t}}_{b}$ as a function of $\kappa\Delta{r}$.
The value is typically ${\hat{t}}_{b}=0.1 -1$ except for 
sharp peaks at $\kappa\Delta{r}\approx n\pi (n=1,2,3,\cdots)$,
which correspond to force-free cases with $K_{0} \to 0$.
It is interesting to compare the numbers ${\hat{t}}_{b}=1.5\times 10^{-3}$ for 
the dipolar magnetic-field extending to exterior vacuum 
for the same $E_{\rm{p}}$ (Paper I).
The breakup time for the confined field was significantly longer. 
This difference is related to the spatial shear distribution driven by the magnetic-field geometry.
This will be discussed in the next subsection.

\begin{figure}\begin{center}
\includegraphics[scale=0.8]{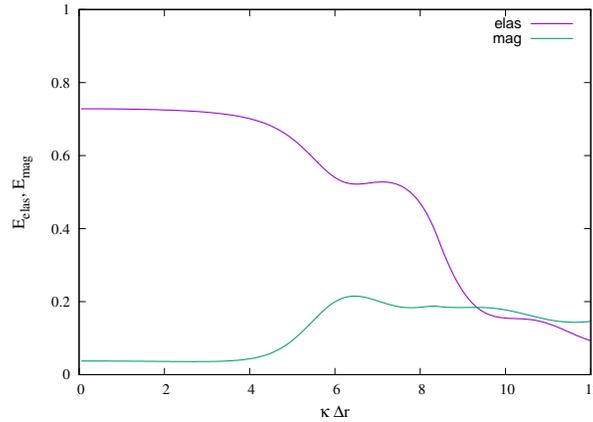}%
\caption{ 
 \label{Fig.co2vsk}
Coefficients ${\hat{E}}_{\rm{elas}}$ 
and ${\hat{E}}_{\rm{mag}}$ as a function of $\kappa \Delta r$.
}
\end{center}\end{figure}

%
 The elastic energy $\Delta E_{\rm elas}$ increased with the square of time $t$.
We numerically integrated over the entire crust and obtained 
\begin{align}
 \Delta E_{\rm{elas}}&=
2\pi\int_{r_{c}} ^{R} r^2dr \int_{0} ^{\pi} \sin\theta d\theta
~\mu\sigma_{ij}\sigma^{ij} 
\nonumber\\
&={\hat{E}}_{\rm{elas}}\times\mu_{1}R^3
\left(\frac{\sigma_{\rm{max}}}{0.1}\right)^2
\left( \frac{t}{t_{*}}\right)^2,
\label{Elsenergy.eqn}
\end{align}
where ${\hat{E}}_{\rm {elas}}$ is the numerical value, 
and $\Delta E_{\rm{elas}}$ was normalized by $\mu_{1}R^3= 5.0\times 10^{45}$~erg 
using $\mu_{1}$ at $\rho_{1}$.
The change in magnetic energy $\Delta E_{\rm{mag}}$ 
associated with $\delta B_{\phi}$ is expressed as 
\begin{align}
\Delta E_{\rm{mag}}&=
2\pi \int_{r_{c}} ^{R} r^2dr \int_{0} ^{\pi} \sin\theta d\theta
~\frac{(\delta B_{\phi})^2 }{8\pi} 
\nonumber\\
&={\hat{E}}_{\rm{mag}}\times E_{\rm{p}}
\left(\frac{\sigma_{\rm{max}}}{0.1}\right)^2\left(\frac{v_{s}}{v_{a}}\right)^4
\left( \frac{t}{t_{*}}\right)^2,
\label{Magenergy.eqn} 
\end{align}
where ${\hat{E}}_{\rm{mag}}$ is the numerical value.
The poloidal magnetic energy 
$E_{\rm p}= 4\times 10^{46} (B_{0}/10^{14} {\rm G})^{2}$~erg
was chosen as the normalization, and other factors were derived using $t_{*}$.
Both ${\hat{E}}_{\rm{elas}}$ and ${\hat{E}}_{\rm{mag}}$ are shown in
 Figure~\ref{Fig.co2vsk} as a function of $\kappa\Delta{r}$.
The numerical factor ${\hat{E}}_{\rm{elas}}$ decreased, whereas
${\hat{E}}_{\rm{mag}}$ increased.
The change with respect to $\kappa\Delta{r}$ was not large and 
${\hat{E}}_{\rm{elas}}$ and ${\hat{E}}_{\rm{mag}}$ were ${\mathcal{O}}(1)$ for all models.
Numerical coefficients in front of 
Equations~(\ref{limitVA.eqn}),(\ref{Elsenergy.eqn}), and 
(\ref{Magenergy.eqn})
and $t_{\rm{d}}/\tau_{\rm{Ohm}}$
are summarized in Table~\ref{table1:mylabel}.
These numbers were also compared with those in Paper I.
A significant difference was observed in the magnetic configuration.
The breakup time for the confined model was typically $10^2$ times longer
than that for the extended model.
This longer timescale led to higher energies
${\hat{E}}_{\rm{elas}}$ and ${\hat{E}}_{\rm{mag}}$,
which typically increased by a factor of $10^4$ 
corresponding to the square of accumulation time.
However, longer timescale constrained the
epoch or magnetic field strength because
the ohmic decay was neglected.
The condition under which $t_{*}\le t_{d}$ is
$B_{0}\ge 3.2\times 10^{13}(\sigma_{\rm{max}}/0.1)^{1/3}$~G
for the extended model.
However, lower field strength
increased by a factor of 5 for
the confined model;
$B_{0}\ge 1.6\times 10^{14}(\sigma_{\rm{max}}/0.1)^{1/3}$~G at least.
%

\begin{table}
    \centering
    \begin{tabular}{cccccc}
Model & Figure &$t_{\rm{d}}/\tau_{\rm{Ohm}}$ 
& ${\hat{t}}_b~~(t_{*})$ & ${\hat{E}}_{\rm elas}$ & ${\hat{E}}_{\rm mag}$\\ 
  \hline \hline
 Extending to exterior & Figure 1-a & $5.9\times 10^{-2}$ &
 $1.5 \times 10^{-3}$ &
 $2.0 \times 10^{-4}$&
 $5.3\times 10^{-5}$
 \\
  Confined to interior and Figure 1-c &
  $3\times 10^{-3}-3\times 10^{-2}$& $0.1-10$ &
 $0.1-0.8$&
 $0.1-0.2$
 \\
    \hline
    \end{tabular}
    \caption{Numerical values 
    for magnetic decay time (\ref{tdiss.eqn}), 
    breakup time(\ref{limitVA.eqn}), elastic energy(\ref{Elsenergy.eqn}),
and magnetic energy(\ref{Magenergy.eqn})}
\label{table1:mylabel}
\end{table}

\subsection{Spatial Shear--Distribution}

\begin{figure}[ht]\begin{center}
\includegraphics[scale=1.2]{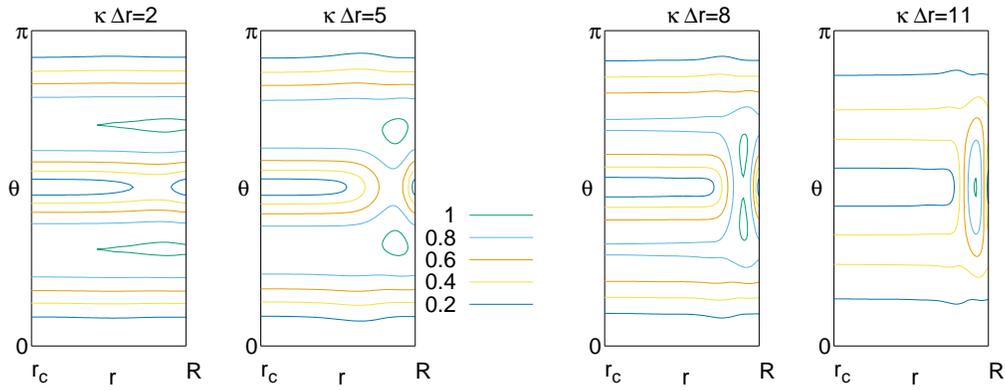}%
\caption{ 
 \label{Fig.shear4}
Contour map of the magnitude $\sigma$ of shear stress normalized by the maximum value.
The left to right panels show the results for 
the models with $\kappa\Delta{r}=$2,5,8, and 11.
}
\end{center}\end{figure}

Figure~\ref{Fig.shear4} shows the magnitude of shear
$\sigma\equiv(\sigma_{ij}\sigma^{ij}/2)^{1/2}$ in the crust.
We found that the shear associated with the axial displacement $\xi_{\phi}$
was significantly larger than that of the polar displacement ${\VEC{\xi}}_{p}=(\xi_{r},\xi_{\theta})$.
This is related to the thin crust of thickness $\Delta{r}/R=0.05$;
typically, $|\xi_{p}|\sim \Delta{r}/R \times |\xi_{\phi}|$.
Polar displacement ${\VEC{\xi}}_{p}$ was induced 
only when the initial MHD equilibrium contained the toroidal component of the magnetic field
($b_{l}\ne 0$ for $\kappa \ne 0$ in Equation~(\ref{cfb1and3.eqn})).
However, the shear distribution shown in Figure~\ref{Fig.shear4} was almost the same
when changing $\kappa$, which determined the ratio of the toroidal to 
poloidal components of magnetic field.
The maximum position of $\sigma$ shifted slightly to the outer radius with increasing radial nodes.
The question that arises is where is the origin of the significant difference, for example, in the breakup time
between the extending model and the confined models.
The initial equilibrium for the former considered in Paper I was purely poloidal.
In contrast to the shear distribution shown in Figure~\ref{Fig.shear4}, 
$\sigma$ exhibited a sharp peak at the surface at $\pm\cos^{-1}(1/\sqrt{3})$
for the extended model (see Figure 2 in Paper I).
The peak originated from the acceleration
$a_{\theta}\ne0$ on the surface of the extended model.
In contrast, $a_{\theta}=0$ because of $\Psi=0$ at the surface
of the confined model (Equation~(\ref{EMacc0.eqn})).
The surface was very fragile because of its weak shear--modulus.
This region was avoided in the confined model, such that
the breakup time to the elastic limit increased.
%

\section{Discussion}

We considered the elastic deformation induced by the evolution of a magnetic field.
The effect of magnetic field geometry was studied and compared with the results of Paper I.
When the field was confined to the interior,
the breakup time for the elastic limit increased by a factor of $10^2$
compared with a field extending to the exterior with the same magnetic energy. 
Accordingly, a larger elastic energy was deposited until crustal fracture.
The elastic energy was typically $\sim 10^{45}$~erg.
The accumulated energy was dependent on the magnetic field geometry and independent of the strength $B_{0}$.
Breakup time $t_{*}$ was proportional to $B_{0}^{-3}$;
$t_{*}\sim 10^{5}(\sigma_{\rm{max}}/0.1)(B_{0}/(10^{14}{\rm{G}}))^{-3}$ years.
As the Ohmic decay of the magnetic field was neglected,
our result is valid for stronger field,
$B_{0}\ge 2\times 10^{14}(\sigma_{\rm{max}}/0.1)^{1/3}$ G at least.
Further, the average strength exceeded  
$2\times 10^{15}(\sigma_{\rm{max}}/0.1)^{1/3}$ G.
The magnetic energy in the crust exceeded $2\times 10^{47}$ erg and the breakup time 
corresponding to the minimum strength was $\sim 10^4$ year.
Unless the field strength $B_{0}$ was significantly larger than $2\times 10^{14}$ G,
the elastic deformation did not reach the critical limit. 
The magnetic field in CCOs may be stable and
 gradually decays owing to the Joule loss.
Our magnetic configuration is limited to a simple case
that is, it has a dipolar angular configuration and a few radial nodes.
The elastic limit in a more general configuration is worth discussing 
based on the results of this study
because a realistic magnetic field in CCOs is more complicated.
When the number of nodes increases in either the angular or radial direction,
spatial size around the maximum shear strain decreases. 
The elastic energy was deposited until the critical limit decreased.
Assuming that the accumulated elastic energy is released during an outburst, such an event is less energetic.
Thus, the number of radial nodes may be effective, 
because the stellar structure changed significantly in the
radial direction, and the outer part near the surface was more prone to breaking.
Simultaneously, ohmic dissipation became effective near the surface.
Therefore, it would be interesting to investigate whether 
a small-scale irregularity in the magnetic field leads to
elastic limit or decay.
However, in the highly tangled limit, 
the magnetic field was irregular on a small scale
and the direction was random.
Thus, the magnetic force can be regarded as 
isotropic magnetic pressure, which causes irrotational force.
It is difficult to drive elastic deformation;
thus, the confined field was stable against elastic fractures.
A strong magnetic field in CCOs is hidden in the crust and is unlikely to lead to
outbursts that occur in magnetars, although the field strength in both sources is of the same order.
The field geometry exhibits a remarkable difference.
Observations of burst events were not reported, except for CCO at RCW 103.
Further, the central neutron star was classified as a magnetar with spin period
$\sim 6.7$~h~\citep{2016MNRAS.463.2394D}.
Future studies will be conducted to examine a simple idea of different field geometries resulting in the occurrence or absence of outburst in strongly magnetized neutron stars.
%

\section*{Acknowledgements}
 This work was supported by JSPS KAKENHI Grant Numbers 
 JP17H06361, JP19K03850(YK),
 JP19K14712, JP21H01078, JP22H01267, JP22K03681(SK), JP20H04728(KF).

 \bibliography{kojima23Jan}
\bibliographystyle{aasjournal} 

    \end{document}